\documentclass[aps,prl,twocolumn,showpacs]{revtex4}

\usepackage{amsmath}
\usepackage{amssymb}
\usepackage{braket}

\begin{document}

\title{Detecting axions via induced electron spin precession} 

\author{Stephon Alexander}
\email{stephon_alexander@brown.edu}
\affiliation{Department of Physics, Brown University, Providence, RI, 02906}
\author{Robert Sims}
\email{robert_sims@brown.edu}
\affiliation{Department of Physics, Brown University, Providence, RI, 02906}

\date{\today }

\begin{abstract}

We propose a new window to detect axion-like particle (ALP) dark matter from electrically charged fermions, such as electrons and quarks.  We specifically consider a direct interaction between the axion and the electron and find that the non-relativistic quantum dynamics induces a spin precession due to the axion and is enhanced by the application of an external electric field.  This precession gives a change in magnetic flux which under certain circumstances can yield a detectable signal for SQUID magnetometers.

\end{abstract}


\maketitle


One of the most compelling dark matter candidates is the Invisible Axion which solves the strong CP problem in QCD \cite{Preskill:1982cy,pierre}.  Furthermore, axion-like particles (ALP), that do not solve the strong CP problem, can also be a viable dark matter candidate \cite{Marsh:2015xka,Dienes:2011ja}.  ALP arise naturally from string theory and are necessary for anomaly cancellation via the Green-Schwarz mechanism \cite{Green:1984sg,FNN}.  Despite these well motivated and important reasons for its role in physics, the Invisible Axion and ALP have evaded detection in both astrophysical and Earth-based experiments.  For example, it is well known that the axion modifies the Maxwell equations by having a new interaction $\delta \cal{L}\rm \propto \phi E\cdot B$ \cite{Preskill:1982cy,Abbott:1982af,Dine:1982ah}.  Some experiments have capitalized on enhancing detection in a resonant cavity with a strong external magnetic field \cite{Sikivie:1983ip,Asztalos:2001tf,Barbieri:2016vwg}.  Others have exploited the possibility of detecting the change in flux from a carefully oriented external magnetic field \cite{Asztalos:2009yp,Kahn:2016aff}.  In this work we consider a new possibility of detecting the axion directly from its interaction with electrons.

A while ago, various authors considered couplings of the axion to matter fields in the standard model \cite{Srednicki:1985xd,Shifman,Zhitnitsky,Kim}.  For the electron, it is possible to have direct couplings and radiatively induced couplings to the axion.   As we will see, the relativistic axion-electron interaction will induce a non-relativistic interaction that involves an axion, electric field, electron coupling which will cause spin precession in the electron wave function; an electric dipole moment.   This effect is similar to how a spin-magnetic field coupling can lead to spin precession.  Similar mechanisms have been considered \cite{Hill:2015kva, Cao:2017ocv} and find similar forms for an induced electron electric dipole moment.  In this letter, we consider new interactions and the quantum mechanics of electrons in the presence of axion dark matter and an external electric field.  We will find that there is an induced change in magnetic flux that is in principle detectable for realistic background field values.  Finally, we propose an idealized experiment, similar to \cite{Budker:2013hfa}, which may detect such a change in flux \cite{FN1}.


Consider dimension-four operators coupling a U(1) gauge field $A_{\mu}$, fermion $\Psi$, and real pseudoscalar $\phi$ that retain gauge invariance and shift-symmetry.  A simple example, analogous to the simplest realizations of Invisible Axion scenarios, contains an extra Higgs singlet is introduced whose phase is the axion $\Phi = 
\rho e^{i\phi/f}$.  Yukawa couplings to quarks and leptons yield the following shift symmetric axion couplings:
\begin{align}
\mathcal{L}_\phi = &-\frac{1}{2}\partial_{\mu}\phi \partial^{\mu} \phi - \mu^{4} \left[1-\cos\left(\frac{\phi}{f}\right) \right]\nonumber\\
& + \lambda f\sin \left(\frac{\phi}{f}\right)\bar{\Psi}i\gamma^5\Psi, \label{EQ:ModelHELag}
\end{align}
where $\lambda$ is the dimensionless Yukawa coupling of the singlet $\Phi$  and fermions $\Psi$, and $\mu$ is a parameter related to instanton effects.  For a detailed description of low energy fermionic coupings for the QCD axion, see \cite{Georgi:1986df}.

In this work we will be studying ultra light axion dark matter solutions, given in Eq.~(\ref{axform}),  where $\phi \ll f $.  This will reflect a symmetry breaking, where the axion acquires a mass by settling into one of the degenerate minima of its (effective) cosine potential.  Without loss of generality, we assume the axion settles into $\phi = 0$ minima, and the small field expansion for $\phi/f$ is applied to the Lagrangian.

The resulting effective Lagrangian can be written as
\begin{align}
\mathcal{L} = \bar{\Psi}\left(i\gamma^{\mu}D_{\mu} - m\right)\Psi -i\lambda\phi\bar{\Psi}\gamma^5\Psi + \mathcal{L}_{\text{kin}}
\end{align}
where $D_{\mu}$ is the U(1) gauge covariant derivative, and $\mathcal{L}_{\text{kin}}$ contains the kinetic terms for the pseudoscalar and the gauge field.  In particular, we consider interactions between electromagnetism, electrons, and the axion.  The equation of motion for the fermion field $\Psi$ is found as
\begin{equation}
\left(i\gamma^\mu\partial_\mu - m + g\gamma^\mu A_\mu - i\lambda\phi\gamma^5\right)\Psi = 0.
\end{equation}
We want to find a non-relativistic form of the equation of motion, analogous to the Schrodinger equation.

Working in the Dirac basis, define $A_0 = \varphi$ and decompose the Dirac fermion four-spinor $\Psi$ into two component spinors.  Then, the equation of motion gives coupled differential equations for the two-component spinors
\begin{align}
\left(E+g\varphi-m\right)\Psi_e = -\left(-i\lambda\phi +\vec{\sigma}\cdot\left(\vec{p}+g\vec{A}\right)\right)\Psi_{\bar{e}},\\
\left(E+g\varphi+m\right)\Psi_{\bar{e}} = \left(-i\lambda\phi - \vec{\sigma}\cdot\left(\vec{p}+g\vec{A}\right)\right)\Psi_e.
\end{align}
In taking the non-relativistic limit, the limit $g\varphi \ll m$ is imposed, as well as the usual approximation $E\approx m$.  Taking these approximations, the equation for $\Psi_{\bar{e}}$ becomes
\begin{equation}
2m\Psi_{\bar{e}} \approx -\left(i\lambda\phi+\vec{\sigma}\cdot\vec{\pi}\right)\Psi_e, \label{PositronEQ}
\end{equation}
where we have defined $\vec{\pi} = \vec{p}+g\vec{A}$. For $\lambda\phi\ll m$, the amplitude of the positron $\Psi_{\bar{e}}$ is suppressed when compared to the electron's amplitude $\Psi_e$.  This condition naturally arises due to the small coupling between the axion dark matter and standard model fermions, and the small expectation value for the axion due to symmetry breaking.

After redefining the energy as the non-relativistic energy $E\rightarrow E+m$, solving for $\Psi_{\bar{e}}$ gives the uncoupled equation for $\Psi_e$ can be found.  Given that the fermion mass is the largest parameter in the problem, we expand the equation of motion in orders of $1/m$.  To lowest order, the non-relativistic equation of motion is
\begin{align}
E\Psi_e &= \left[\frac{1}{2m}\left(i\vec{\nabla}+g\vec{A}\right)^2 + 2\left(\frac{g}{2m}\right)\vec{S}\cdot\vec{B} - g\varphi\right]\Psi_e\nonumber\\
&+ 2\left(\frac{g}{2m}\right)\left[\vec{S}\cdot\vec{\nabla}\left(\frac{\lambda}{g}\phi\right)\right]\Psi_e +\frac{\left(\lambda\phi\right)^2}{2m}\Psi_e.
\end{align}
Written this way, it appears that spatial gradients of the axion field can act as an effective magnetic field for the electrons with value $\vec{B}_{\text{eff}} = \frac{\lambda}{g}\vec{\nabla}\phi$.

Inclusion of the next order corrections introduces many important phenomena to the quantum mechanical description of the electron, including the spin-orbit coupling.    Additionally, terms will appear in the non-relativistic Hamiltonian for an electron interacting with electromagnetic fields and axions.  In particular, the new axion interaction terms, to second order, are given by
\begin{equation}
H_{axion} = \frac{\lambda}{m}\left(1-\frac{g\varphi}{2m}\right)\left[\vec{S}\cdot\vec{\nabla}\phi +\frac{1}{2}\lambda\phi^2\right]+\left(\frac{g\lambda\phi}{2m^2}\right)\vec{S}\cdot\vec{\mathcal{E}},\label{FullHAx}
\end{equation}
where the electric field is defined as $\vec{\mathcal{E}}=-\vec{\nabla}\varphi$.

We want to understand which is the dominant term.  In the non-relavistic regime, $g\varphi \ll m$, hence the first term in $H_{axion}$ can be looked at as simply the $1/m$ dependence.  In other words, we want to compare the magnitudes of $\lambda\vec{\nabla}\phi$  and $\frac{g\lambda}{m}\phi\vec{\mathcal{E}}$.  The first term is a quantity set by the axion field, which we cannot control.  However, the second term depends on the external electric field.  Hence, we want to find some condition on the electric field magnitude.  We do not consider the $\lambda^2\phi^2$ term as it will only produce a uniform shift in the energy of the electron.

Consider a model where the axion $\phi$ is the principal component of our local dark matter energy density $\rho_{DM}$.  We approximate the axion field, as in \cite{Arvanitaki:2015iga}, by
\begin{equation}
\phi(t,x) \approx \frac{\sqrt{2\rho_{DM}}}{m_\phi}\cos\left[m_\phi\left(t-\vec{v}\cdot\vec{x}\right)\right]\label{axform}
\end{equation}
where $m_\phi$ is the axion mass, $\vec{v}$ is the virial velocity in our galaxy $|\vec{v}|\sim 10^{-3}$.  Then, the critical value of the electric field is
\begin{equation}
\mathcal{E} = \frac{m_e}{g}\frac{\nabla\phi}{\phi}\sim (3\times 10^{9} \text{ V/m})\left(\frac{m_\phi}{1\text{ eV}}\right).\label{Ethresh}
\end{equation}
Most dark matter model use values of $m_\phi \leq 10^{-6}$ eV, giving the $\mathcal{E} \sim 1$ kV/m.  Above this value, the $\vec{S}\cdot\vec{\mathcal{E}}$ term is the dominant axion-electron interaction term.  For the remainder of the calculation, we assume that we are above this critical electric field and consider only the additional term
\begin{equation}
H_{axion} = \left(\frac{g\lambda\phi}{2m_e^2}\right)\vec{S}\cdot\vec{\mathcal{E}}.\label{AxHam}
\end{equation}


For now, we consider a what happens for a single electron subject to electromagnetic fields.  The axion field term is considered to be a perturbative addition to the Hamiltonian, $H_1$.  We ignore the spin-orbit coupling term for simplicity, however in the presence of a magnetic field,  we expect this term will be at least as important as the axion correction term.  Furthermore, we wish to isolate the effects of the new axion interaction by explicitly setting the magnetic field to zero.  Written explicitly, we consider the following Hamiltonian for the electron:
\begin{equation}
H = -\frac{1}{2m_e}\nabla^2 - g\varphi + \frac{g\lambda}{2m_e^2}\vec{S}\cdot\vec{\mathcal{E}}\;\phi(\vec{x},t).
\end{equation}
We take constant electric field $\vec{\mathcal{E}} = \mathcal{E}\hat{z}$.  We want to find the commutator
\begin{equation}
\left[\frac{p^2}{2m_e} -g\varphi, \frac{g\lambda\mathcal{E}}{2m_e^2}S_z\phi(\vec{x},t)\right] = \frac{g\lambda\mathcal{E}}{4m_e^3}S_z\left[p^2,\phi(\vec{x},t)\right]
\end{equation}
where we note the Hilbert space associated with the spin is disjoint from the spatial dependence.  The remaining commutator is in general nonzero.  Using the form of the axion field from Eq.~(\ref{axform}), each spatial gradient of $\phi$ is suppressed by a factor $m_\phi v \ll 10^{-8}$ eV.  As a result, we approximate the axion field as spatially homogeneous $\phi(\vec{x},t) = \langle\phi\rangle$, giving $\left[p^2,\phi(\vec{x},t)\right] = 0$.  Therefore, we use a basis that simultaneously diagonalizes the Hamiltonian, $\psi_n(\vec{x})\ket{\pm}$ defined by
\begin{gather}
H_0\psi_n(\vec{x},t) = E_n \psi_n(\vec{x},t),\\
H_1\ket{\pm} = \pm\frac{g\lambda\mathcal{E}}{4m_e^2}\langle\phi\rangle\ket{\pm}.
\end{gather}
The axion interaction Hamiltonian results in splitting in the electron energy spectrum.  As an example, consider some initial state
\begin{equation}
\Psi(\vec{x}, t=0) = \psi_n(\vec{x},0)\left(\frac{\ket{+}+\ket{-}}{\sqrt{2}}\right)
\end{equation}
such that $\int|\psi_n(\vec{x}, 0)|^2 =1$.  The expectation values of spins in each direction at some later time $t$ is given by
\begin{gather}
\langle S_x\rangle = \frac{1}{2}\cos\left(\frac{g\lambda\mathcal{E}\langle\phi\rangle}{2m_e^2}t\right), \;\;\;
\langle S_y\rangle = \frac{1}{2}\sin\left(\frac{g\lambda\mathcal{E}\langle\phi\rangle}{2m_e^2}t\right),\nonumber\\
\langle S_z\rangle = 0.
\end{gather}
We recognize this as a spin precession phenomena where the electric field is aligned in the $\hat{z}$-direction and the initial configuration of spins is in the $\hat{x}$-direction.  The timescale for this spin precession, using the threshold electric field, given by Eq.~(\ref{Ethresh}), and local dark matter energy $\rho_{DM} \sim 0.3$ GeV/cm$^3$, is
\begin{equation}
\tau = \frac{2m_e^2}{g\lambda\mathcal{E}\langle\phi\rangle}\sim \frac{2m_e^2}{g\mathcal{E}}\left(\frac{m_\phi}{\lambda\sqrt{2\rho_{DM}}}\right) \sim \frac{10^{-4} s}{\lambda}.
\end{equation}
Note, for a given constant electric field strength, there is still a linear dependence on the mass of the axion.  The lighter the axion, the larger this effect should be.

Consider now a collection of $N$ electrons all prepared in the $+\hat{x}$-direction,  as the single electron case.  We expect the coupling $\lambda$ between the axion and electrons is small, then the timescale for the precession is large.  The magnetic field in the $\hat{x}$-direction varies inversely to the square of the timescale, and thus is treated as constant.  However, in the $\hat{y}$-direction, the magnetic moment of the electrons is
\begin{equation}
\mu_{y} \sim 2\mu_B N\langle S_y\rangle = \mu_B N\sin \left(\frac{t}{\tau}\right).
\end{equation}
where $\mu_B$ is the Bohr magneton.  We now imagine a loop of wire whose norm is in the $\hat{y}$-direction.  If the loop is taken to be the same size as the collection of electrons with cross section $A$, then the magnetic flux through the loop of wire will be
\begin{equation}
\Phi_B(t) \sim \mu_B \mu_0 nA\sin\left(\frac{t}{\tau}\right)\label{Flux}
\end{equation}
with $n$ number density of electrons.  For non-interacting electrons, we must ensure the deBroglie wavelength is larger than the average distance between electrons.  In particular, $n\lambda_{dB}^3 < 1$.  Saturating the inequality gives a maximum number density allowed.  At some small time $t$ relative to the timescale $\tau$, the rate of change of flux is
\begin{equation}
\frac{d\Phi_B}{dt}\bigg|_{t=0} = \frac{e\mu_B \mu_0}{2m_e^2}\left(\frac{\lambda\sqrt{2\rho_{DM}}}{m_\phi}\right)nA\cdot\mathcal{E}.\label{FluxChange}
\end{equation}
The changing flux will be inversely proportional to the timescale $\tau$.  We also assume that the electric field will not change the cross-sectional area of the collection of electrons.  Dissipation of the electrons may provide an experimental problem.  However, in the regime where the dissipation rate satisfies
\begin{equation}
\frac{dA}{dt} \ll \frac{A}{t},
\end{equation}
the flux change due to a decrease in number density is a subleading effect.


Including the axion-electron interaction results in a classical electric dipole moment for the electron, as seen in Eq.~(\ref{AxHam}).  In general, an electric dipole term can be written in the form \cite{Pospelov:2005pr}
\begin{equation}
H = \frac{d_e}{S} \vec{S} \cdot \vec{\mathcal{E}}.
\end{equation}
The electric dipole moment induced by the axion can be found, by comparison, as
\begin{equation}
d_e = \frac{e\lambda}{m_e^2}\frac{\sqrt{2\rho_{DM}}}{m_\phi}\cos(m_\phi t) \label{EQ:EDM}
\end{equation}
While the Standard Model predicts a nonzero electron electric dipole moment due to loop correction, the current experimental bound is given $d_e\leq 8.7\times10^{-29} e\cdot$cm \cite{Baron:2013eja}.  Converting this bound to one on the parameters $\lambda, m_\phi$ gives
\begin{equation}
\lambda \left(\frac{1\text{ eV}}{m_\phi}\right)\leq 10^{-10}.
\end{equation}
Saturating the bound, the change in flux given by Eq.~(\ref{FluxChange}) can be found as
\begin{equation}
\frac{d\Phi_B}{dt} \sim 10^{-18}\text{ Wb}/s \label{dFlux}
\end{equation}
for number density $n\sim 10^{21}$ m$^{-3}$, electric field $\mathcal{E} \sim 10^5$ V/m, and cross sectional area $A\sim 1$ m$^2$.

The frequency of oscillation of the electric dipole moment from Eq.~(\ref{EQ:EDM}) matches the previous results of \cite{Hill:2015kva}, which rely on different interaction terms.  This frequency is a universal feature of treating the axion as a classical oscillating field.  In our analysis, however, we treat the pseudoscalar Yukawa interaction as a necessary term in the effective field theory.  For axion models solving the strong CP problem, \cite{Georgi:1986df} provides a comprehensive analysis for finding the low energy interactions of the axion, including the particular value of $\lambda$.

More generally, the dimensionless coupling constant $\lambda$ is determined by the particular ALP model.  For example in the DFSZ model,
\begin{equation}
\lambda = \frac{m_{e}v^{2}_{u}}{N_{DW} f_{a}v^{2}_{EW}}.
\end{equation}
In string theory, where there are many axions $\lambda = C_{ie}m_{e}/f_{ai}$ where the index $i$ denotes the number of axions \cite{Cicoli:2012sz}.  In models of many axions this coupling could be larger than models of only one axion \cite{Dienes:2011ja}.


A CP conserving interaction between the axion and electrons contributes multiple axion correction terms to the non-relativistic electron Hamiltonian.   The prominent feature found is the emergence of a spin-electric field coupling that depends on the magnitude of the axion field.  Contrary to other axion couplings, the presence of an interaction absent of derivatives proves robust against a wide range of axion masses.  In particular, if the axion is a major component of the local dark matter energy density, experiments looking for axion-electron interactions can probe the lower spectrum of axion masses.

The dominant correction to the non-relativistic electron Hamiltonian, given by Eq.~(\ref{AxHam}), will result in a classical electric dipole moment.  When subject to an external electric field, the dipole will exhibit spin precession.  For reasonable values of physical parameters, the induced changing magnetic flux can be the same order as the sensitivity of SQUID magnetometers.  Experiments measuring electron electric dipole moment, such as \cite{Baron:2013eja}, use methods with heavy molecules to cause spin precession in the presence of both electric and magnetic fields.  However, these experiments measure fluorescence emissions, not a direct detection the flux change due to precession.

We have primarily considered the resulting electron electric dipole moment, however this is not unique to axion-electron interactions.  Many models, including the Standard Model, predict finite electric dipole moments due to quantum effects.  Collider experiments and dark matter direct detection provide relativistic avenues to search for axion interactions.  However, subleading terms in Eq.~(\ref{FullHAx}) provide additional predictions.  In particular, the gradient of the axion need not be as small as previously stated.  In general, the dark matter energy density will have fluctuations, possibly amplified due to an astrophysical production of axions.  These gradient terms can induce additional energy shifts of the electron as well as modifying the path  of cosmic rays.  Such experiments will introduce measurements with different dependencies on the parameters in the theory than what we have presented.

In particular, the axion solution in Eq.~(\ref{axform}) is a background solution for cold dark matter axions, where interactions are treated as negligible perturbations.  Adding the usual axion-photon coupling, 
\begin{equation}
\mathcal{L} \supset \frac{\alpha}{f_a} \phi \vec{\mathcal{E}}\cdot\vec{B},
\end{equation}
with $\alpha$ the fine structure constant and $f_a$ the energy cutoff for the effective field theory, the ambient electric and magnetic fields can induce an additional axion field gradient.  Then the gradient term in the non-relativistic electron Hamiltonian becomes dominant when the ambient magnetic field projected in the direction of the electric field is above the cutoff
\begin{equation}
B_{||} = \left(10^{12} \text{ T}\right)\left(\frac{1\text{ eV}}{m_\phi}\right)\left(\frac{1 \text{ m}}{L}\right)\left(\frac{f_a}{M_{pl}}\right)
\end{equation}
where $M_{pl}$ is the Planck mass and $L$ is the size of the experimental apparatus. For dark matter axion mass at $10^{-6}$ eV and the energy scale $f_a$ for the Pontryagin term is taken to be $10^{15}$ GeV, the threshold magnetic field is $B \sim 10^{15}$  Tesla.  Furthermore, the induced gradient only dominates the when $\vec{E}\cdot\vec{B} \geq 10^{30}$ T$\cdot$V/m.  Instead, we may also consider the situation where the axion field is screened by baryonic matter.  In this case, $\phi \sim 0$ and the electric dipole moment term will be proportional to the perturbation of the axion field.  The dominant term only depends on the electric field, similar to Eq.~(\ref{Ethresh}), differing only in an additional dependence on the size of the experimental apparatus.  For small experimental setups, the electric field can be weaker for the dominant phenomena to be the electric dipole term.  Again, unless the field strengths are large, such a situation will only further suppress the expected phenomena.

To achieve sufficiently large number for the predicted flux change in Eq.~(\ref{dFlux}), as well as suppress external magnetic fields, superconductors may provide a useful test bed for experiments looking for the spin precession because magnetic fields should be suppressed.  However, suppressing external magnetic fields may not be necessary to detect the precession due to the electric dipole moment \cite{FNDif}.  As an idealized example of differential measurement, in the presence of both electric and magnetic fields, the rotation axis for the spin precession is given by the weighted (by dipole moments) average of the magnetic and electric fields.  If these fields are constant and orthogonal to one another, there should be an observed change in flux in the direction of the magnetic field.  This magnetic field can not be attributed to the magnetic spin precession.  The observed flux change in the direction of the magnetic field will have the same magnitude as in Eq.~(\ref{FluxChange}), but it will oscillate with the frequency of the magnetic spin precession.

Because thermal fluctuations can induce a changing flux in the direction of the magnetic field, thermal effects will be important for similar experiments.  Furthermore, a collection of electrons in a mixed state will not produce the desired spin precession.  Therefore, the collection of electrons must be kept at a low temperature.  For finite temperature, the number density of electrons in Eq.~(\ref{Flux}) can be replaced by the net number density of electrons.

We would like to thank Ian Dell'Antonio, Humphrey Maris, David Spergel, and Jim Valles for their useful discussions and comments.

\end{document}